\documentclass[aps,prd,floatfix,twocolumn,amsmath,amssymb,nofootinbib]{revtex4-1}
\usepackage{amsthm,color,latexsym,graphicx,array,multirow,epstopdf}
\usepackage[hyperfootnotes=true]{hyperref}

\newcommand{\gev}{\mathrm{GeV}}
\newcommand{\tev}{\mathrm{TeV}}
\newcommand{\npv}{ N_{\text{PU}} }

\definecolor{darkred}{rgb}{0.5,0.0,0.0}
\definecolor{darkblue}{rgb}{0.0,0.0,0.5}
\definecolor{darkgreen}{rgb}{0.0,0.5,0.0}

\newcommand{\red}{\color{darkred}}
\newcommand{\blue}{\color{darkblue}}
\newcommand{\green}{\color{darkgreen}}

\begin{document}
\title{Jet Cleansing: Pileup Removal at High Luminosity}

\author{David Krohn}
\email{dkrohn@physics.harvard.edu}
\affiliation{Department of Physics, Harvard University, Cambridge MA, 02138 \vspace{-1ex}}
\author{Matthew Low}
\email{mattlow@uchicago.edu}
\affiliation{Kavli Institute for Cosmological Physics and Enrico Fermi Institute, University of Chicago, Chicago IL, 60637 \vspace{0ex}}
\author{Matthew D. Schwartz}
\email{schwartz@physics.harvard.edu}
\affiliation{Department of Physics, Harvard University, Cambridge MA, 02138 \vspace{-1ex}}
\author{Lian-Tao Wang}
\email{liantaow@uchicago.edu}
\affiliation{Kavli Institute for Cosmological Physics and Enrico Fermi Institute, University of Chicago, Chicago IL, 60637 \vspace{0ex}}

\date{\today}

\begin{abstract} 
One of the greatest impediments to extracting useful information from high luminosity hadron-collider data is radiation from secondary collisions ({\it i.e.} pileup) which can overlap with that of the primary interaction.  In this paper we introduce a simple jet-substructure technique termed \emph{cleansing} which can consistently correct for large amounts of pileup in an observable independent way.  Cleansing works at the subjet level, combining tracker and calorimeter-based data to reconstruct the pileup-free primary interaction.  The technique can be used on its own, with various degrees of sophistication, or in concert with jet grooming.  We apply cleansing to both kinematic and jet shape reconstruction, finding in all cases a marked improvement over previous methods both in the correlation of the cleansed data with uncontaminated results and in measures like $S/\sqrt{B}$.  Cleansing should improve the sensitivity of new-physics searches at high luminosity and could also aid in the comparison of precision QCD calculations to collider data.
\end{abstract}

\maketitle

\section{Introduction}
\label{sec:introduction}

Many interesting signatures both in the Standard Model and beyond are seen at the LHC  in hadronic final states.  This has motivated  recent theoretical work in jet substructure, {\it e.g.}~\cite{Gallicchio:2010sw,Almeida:2011aa,Azatov:2013hya,Backovic:2012jj,Chien:2013kca,Cui:2013spa,Curtin:2012rm,Dasgupta:2013ihk,Dasgupta:2013via,Ellis:2012sd,Ellis:2012zp,Ellis:2012sn,Feige:2012vc,Gallicchio:2012ez,Gallicchio:2011xq,Gouzevitch:2013qca,Han:2012cu,Hedri:2013pvl,Hook:2012fd,Kahawala:2013sba,Krohn:2012fg,Soper:2012pb}, much of which has seen quick adoption in the experimental community (for an overview of the field see~\cite{Salam:2009jx,Altheimer:2012mn, Shelton:2013an,Plehn:2011tg}).
One outstanding problem is pileup (PU), defined as overlapping secondary collisions on top of the primary interaction. 
As a rough rule-of-thumb, each pileup vertex contributes around 600 MeV of energy per unit rapidity per unit azimuth~\cite{Cacciari:2008gn,Cacciari:2009dp,Rubin:2010fc} (in contrast, the underlying event
 contributes around 2-3 GeV of energy density.)
Thus, for $\npv\sim100$, levels which will soon be regularly encountered at the LHC, an $R=1.0$ jet might suffer $200$ GeV of contamination!

 \begin{figure}[t]
   \includegraphics[width=0.40\textwidth]{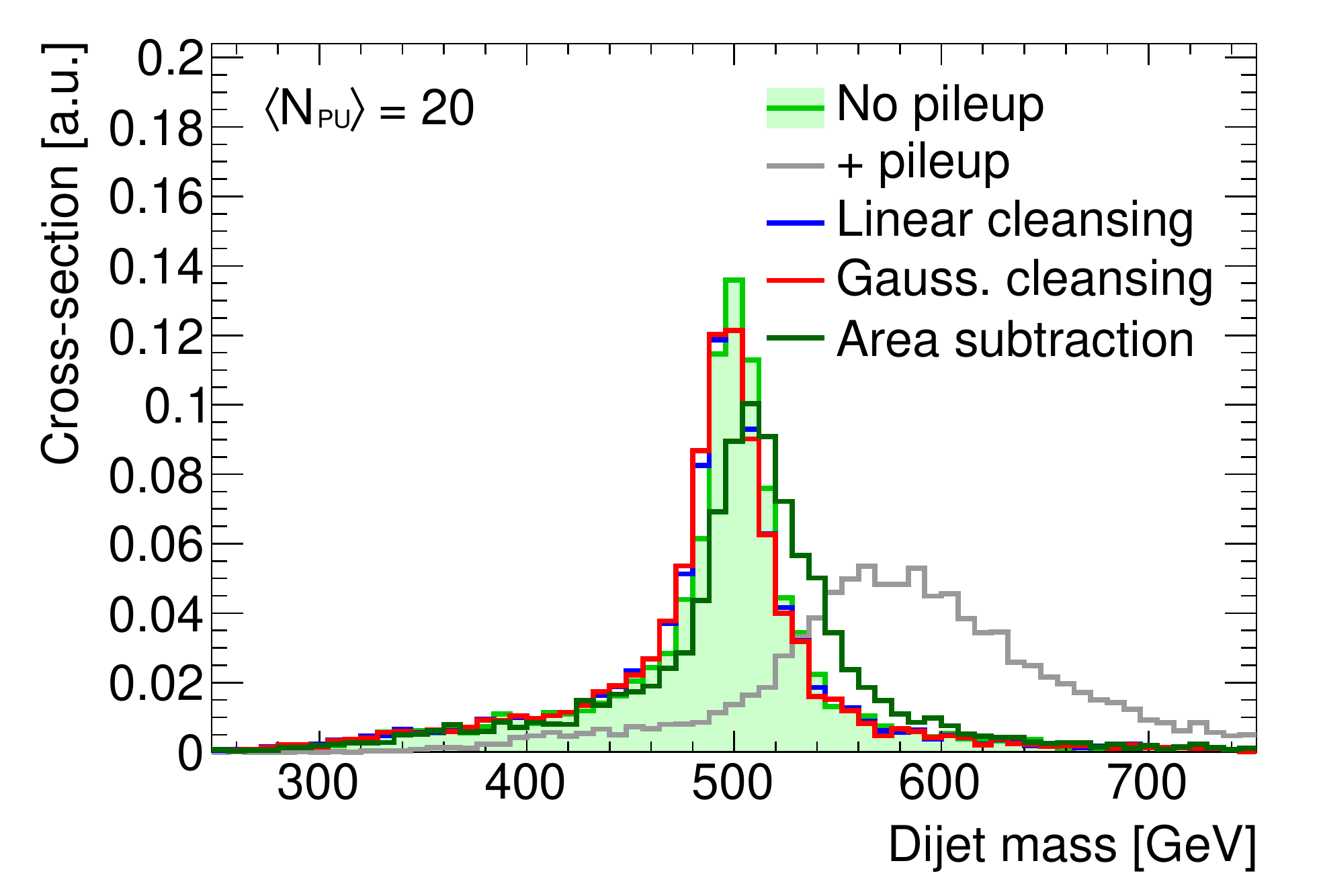}
   \includegraphics[width=0.40\textwidth]{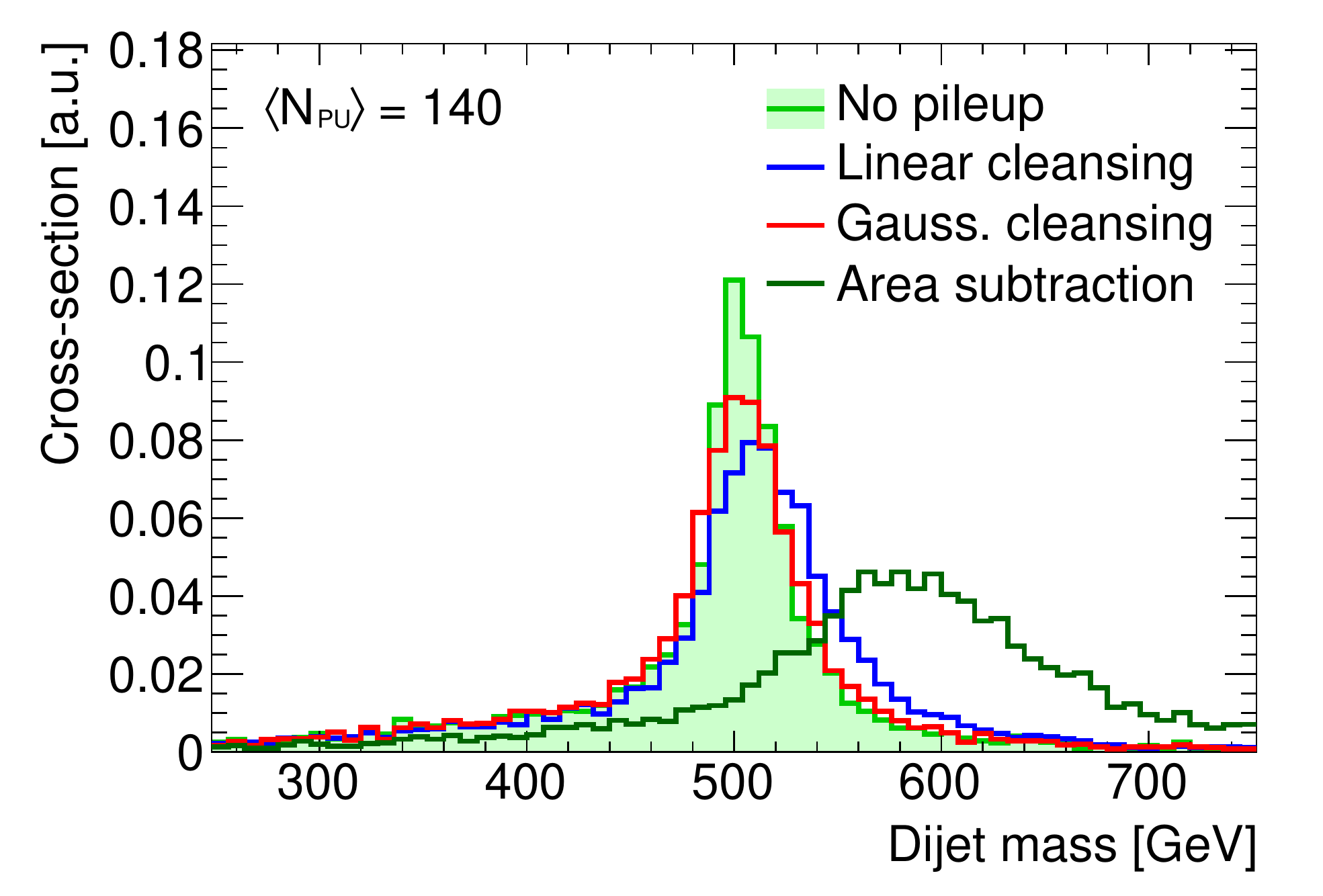}
   \caption{Dijet mass distributions for various methods with 20 and 140 pileup vertices.  Results shown are without grooming, groomed results can be seen in Table~\ref{table:resultsD}.
 \vspace{-3ex}
  \label{fig:dijetmass}
}
 \end{figure}

There are already a number of very effective tools for pileup removal. 
The trackers at both the ATLAS and CMS experiments can determine with excellent accuracy whether a charged particle harder than around 500 MeV came from the leading vertex or a pileup vertex. Thus, most of the  charged hadrons from pileup can be simply discarded -- 
a method called \emph{charged hadron subtraction} (CHS) which is used by CMS. An alternative, popular in the ATLAS collaboration, is to use the \emph{Jet Vertex Fraction} (JVF) -- defined as the fraction of track energy coming from the leading vertex. Cutting on the JVF can effectively remove pileup jets.

Over the last few years, these  solutions have been bolstered by new ideas coming from jet substructure.  These fall into roughly two classes: (1)  \emph{Jet area subtraction}~\cite{Cacciari:2007fd} estimates the amount of pileup in a particular
jet from the pileup density $\rho$ outside of the jet, on an event-by-event basis. Subtracting off $\rho~\times$ area from the jet energy successfully restores distributions of kinematic observables to close to their uncontaminated state.
Through a clever modification called \emph{shape subtraction}
this technique can also be applied to more general jet shapes~\cite{Soyez:2012hv}.  
(2) \emph{Jet grooming} techniques ({\it i.e.} filtering~\cite{Butterworth:2008iy},  pruning~\cite{Ellis:2009su,Ellis:2009me}, and trimming~\cite{Krohn:2009th})  attempt to  identify individual pileup emissions within jets and remove them dynamically.   
Methods from both classes, as well as combinations of methods, have already proven
effective in 7 and 8 TeV LHC data.
  
 \begin{figure}[t]
   \includegraphics[width=0.40\textwidth]{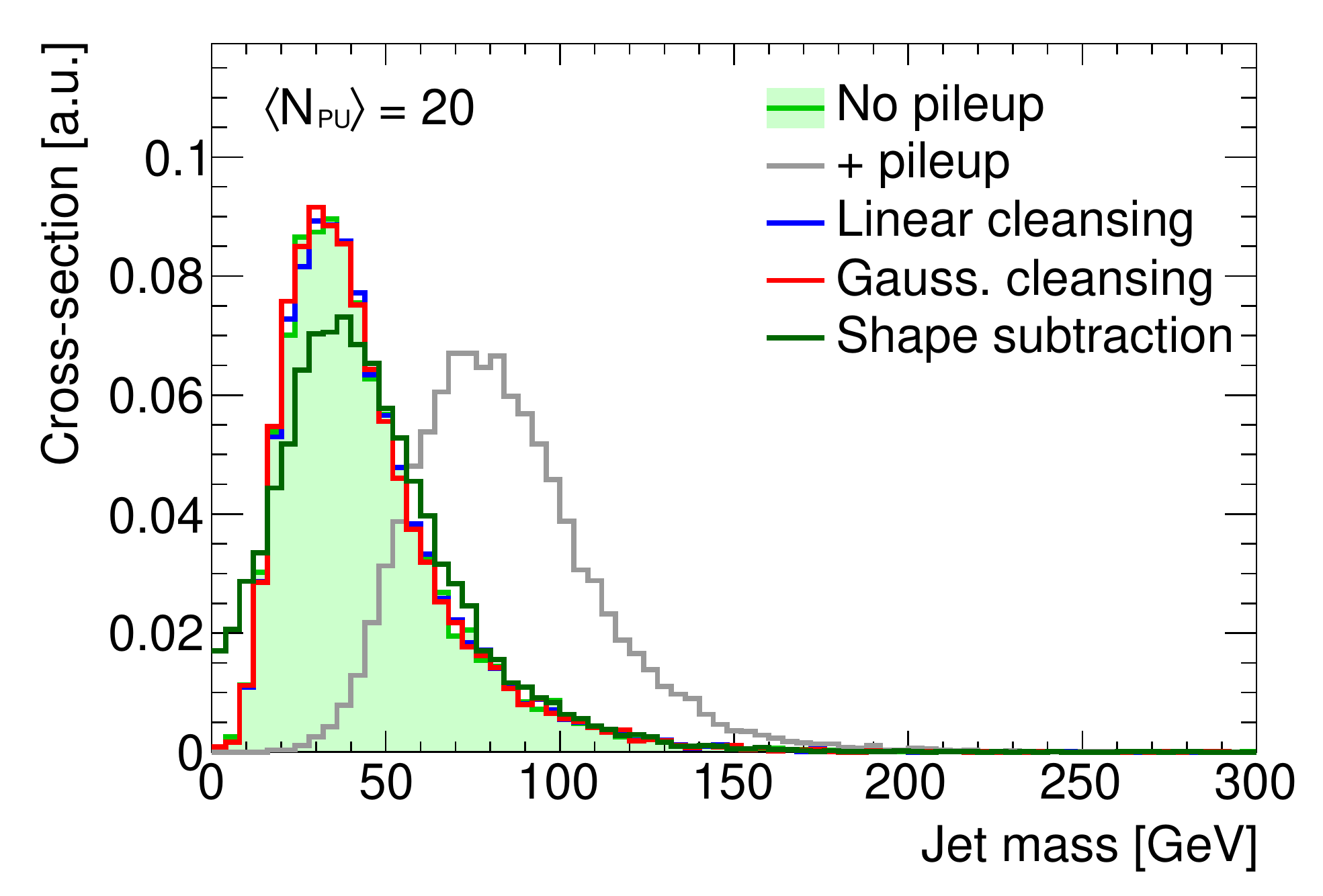}
   \includegraphics[width=0.40\textwidth]{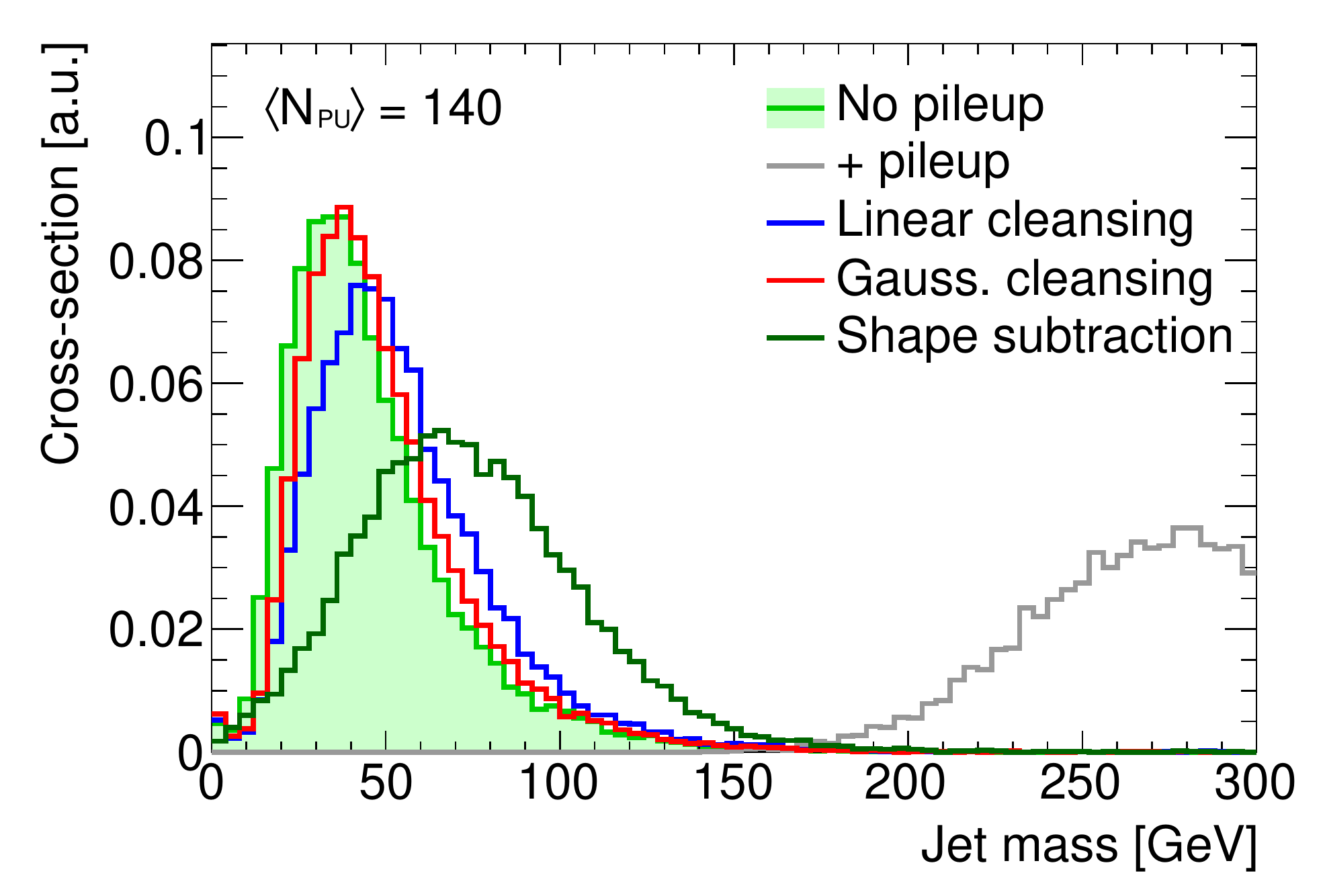}
   \caption{Jet mass distributions for various methods with 20 and 140 pileup vertices.  Results shown are without grooming, groomed results can be seen in Table~\ref{table:resultsD}.
 \vspace{-3ex}
  \label{fig:jetmass}
}
 \end{figure}

\begin{figure}[t]
  \includegraphics[width=0.40\textwidth]{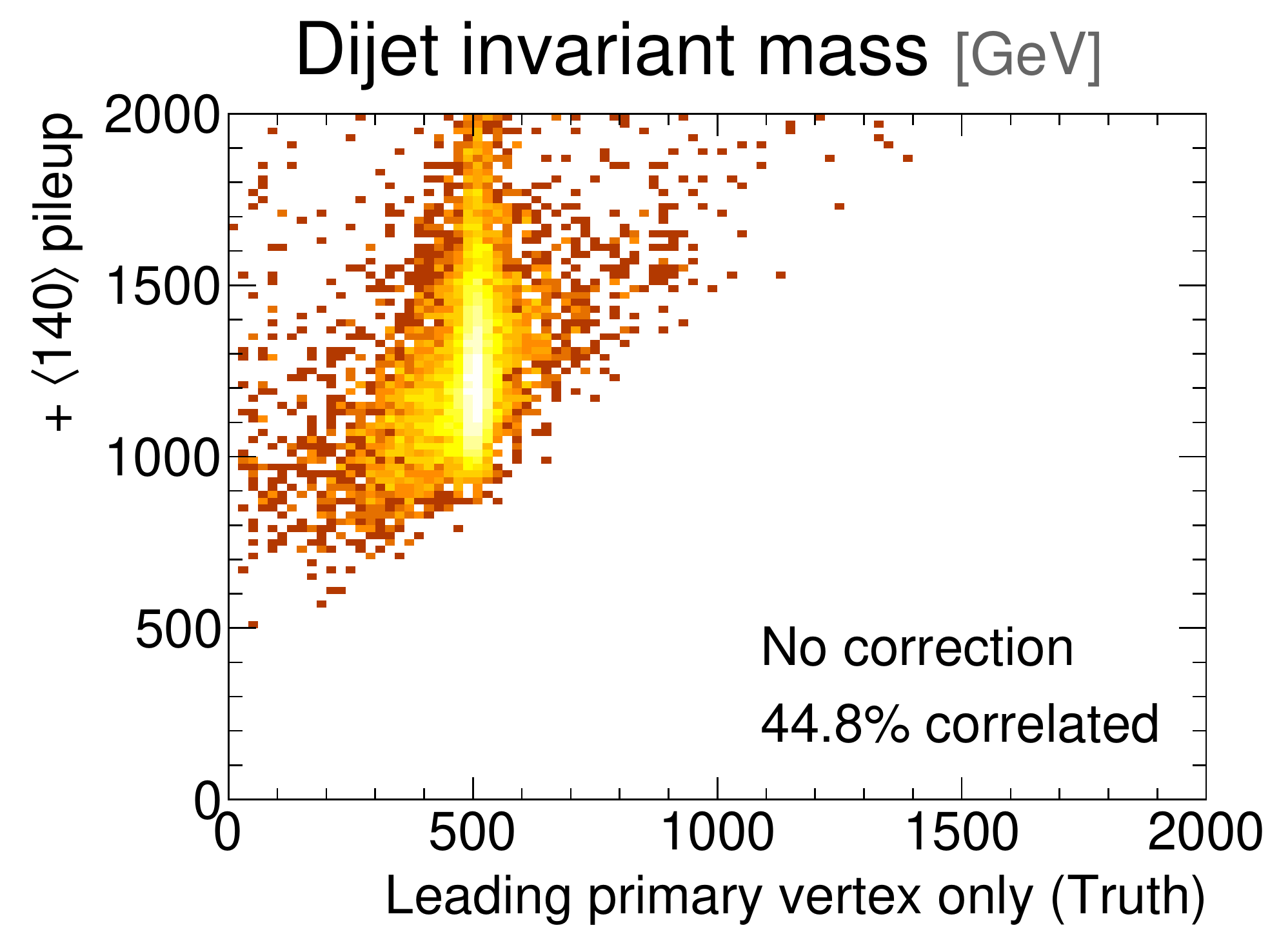}
  \includegraphics[width=0.40\textwidth]{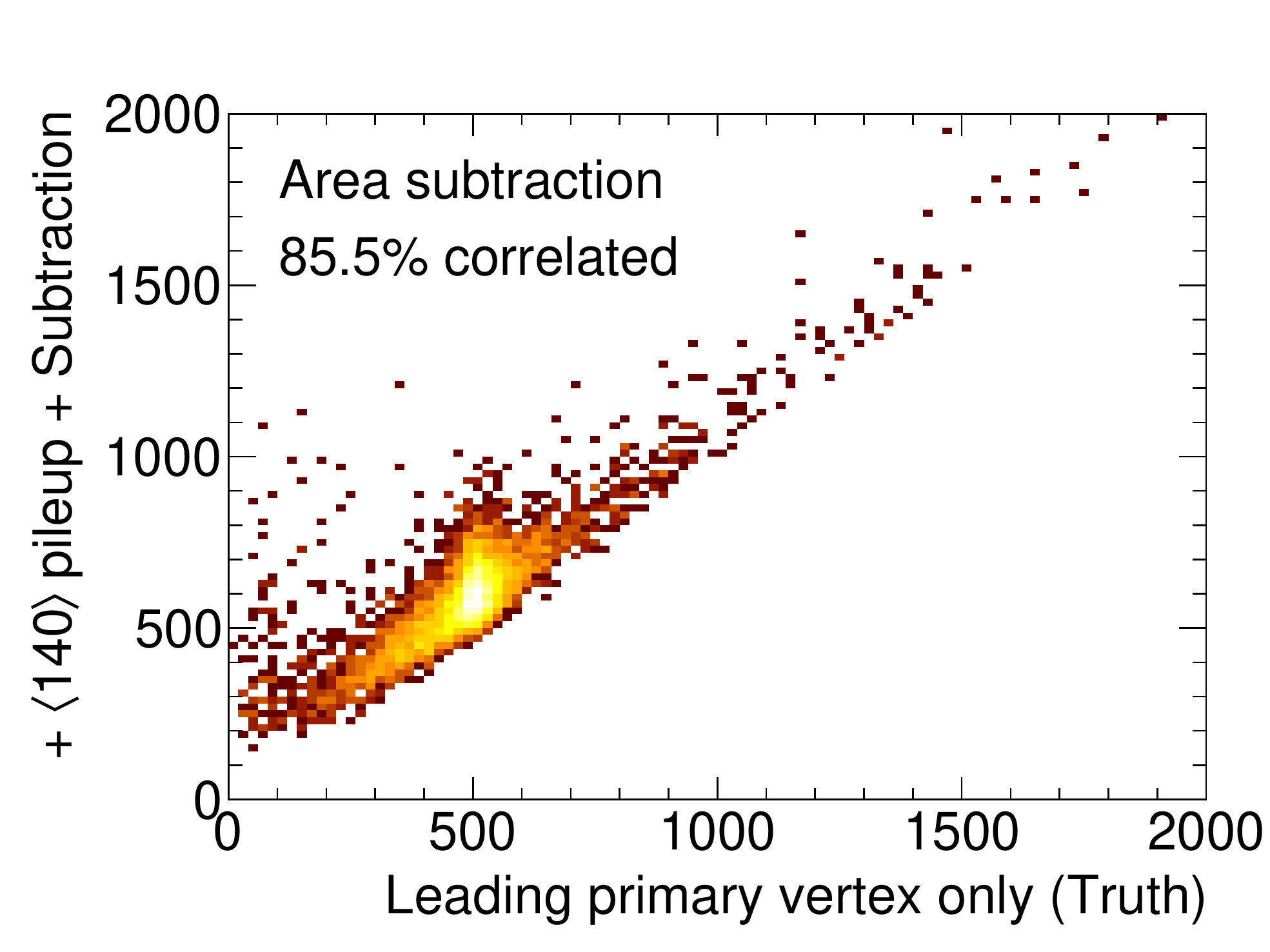}
  \includegraphics[width=0.40\textwidth]{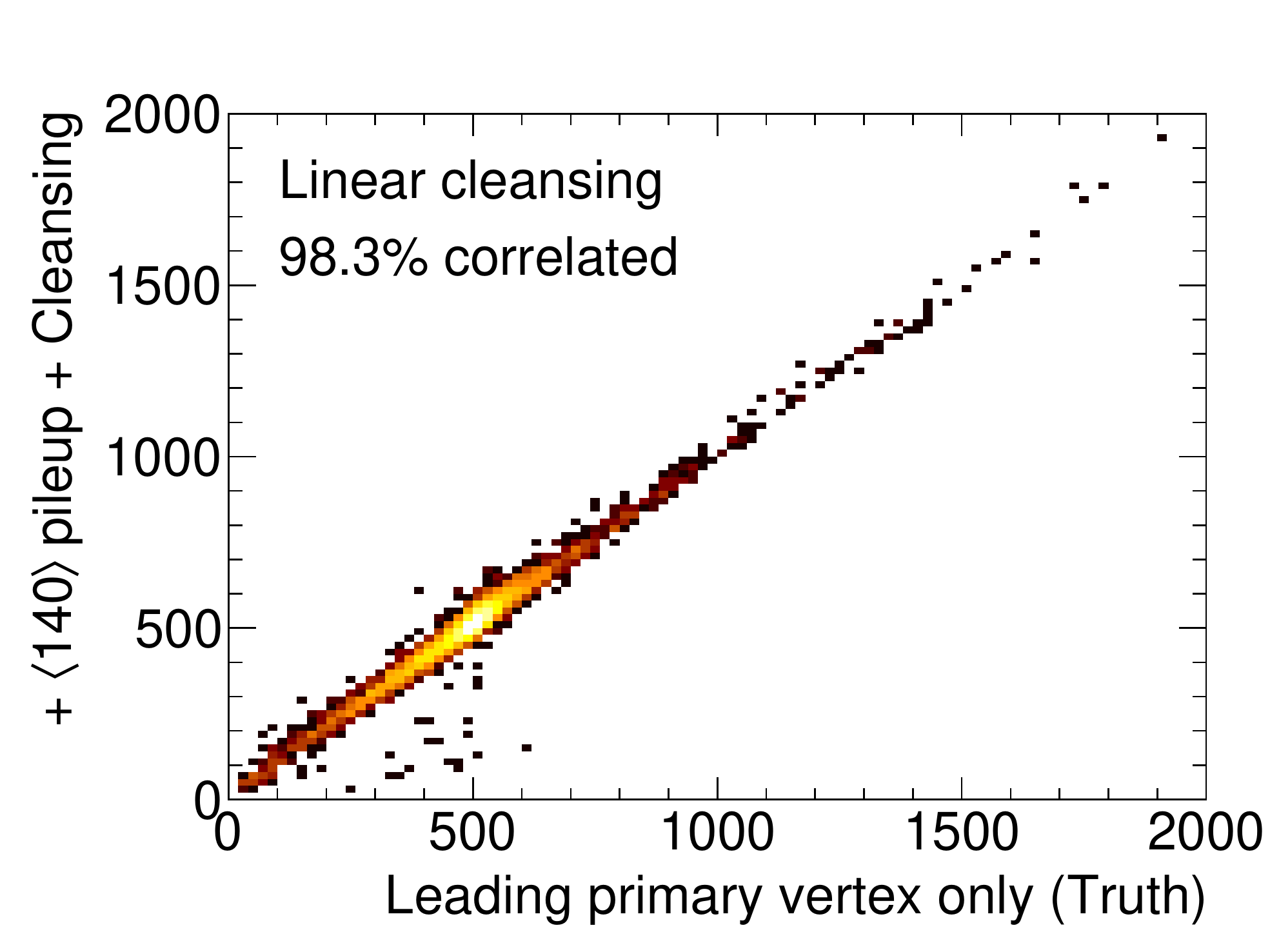}
  \caption{Correlations for dijet mass, a kinematic variable, are shown between between events with 140 pileup interactions, corrected via subtraction or cleansing, and the truth version of the same events, with pileup explicitly removed.  The top row shows the uncorrected correlations, the middle row demonstrates the performance of~\cite{Cacciari:2007fd}, and the bottom row shows the performance of the linear cleansing method described here.}
  \vspace{-3ex}
  \label{fig:dijet2D}
\end{figure}

\begin{figure}[t]
  \includegraphics[width=0.40\textwidth]{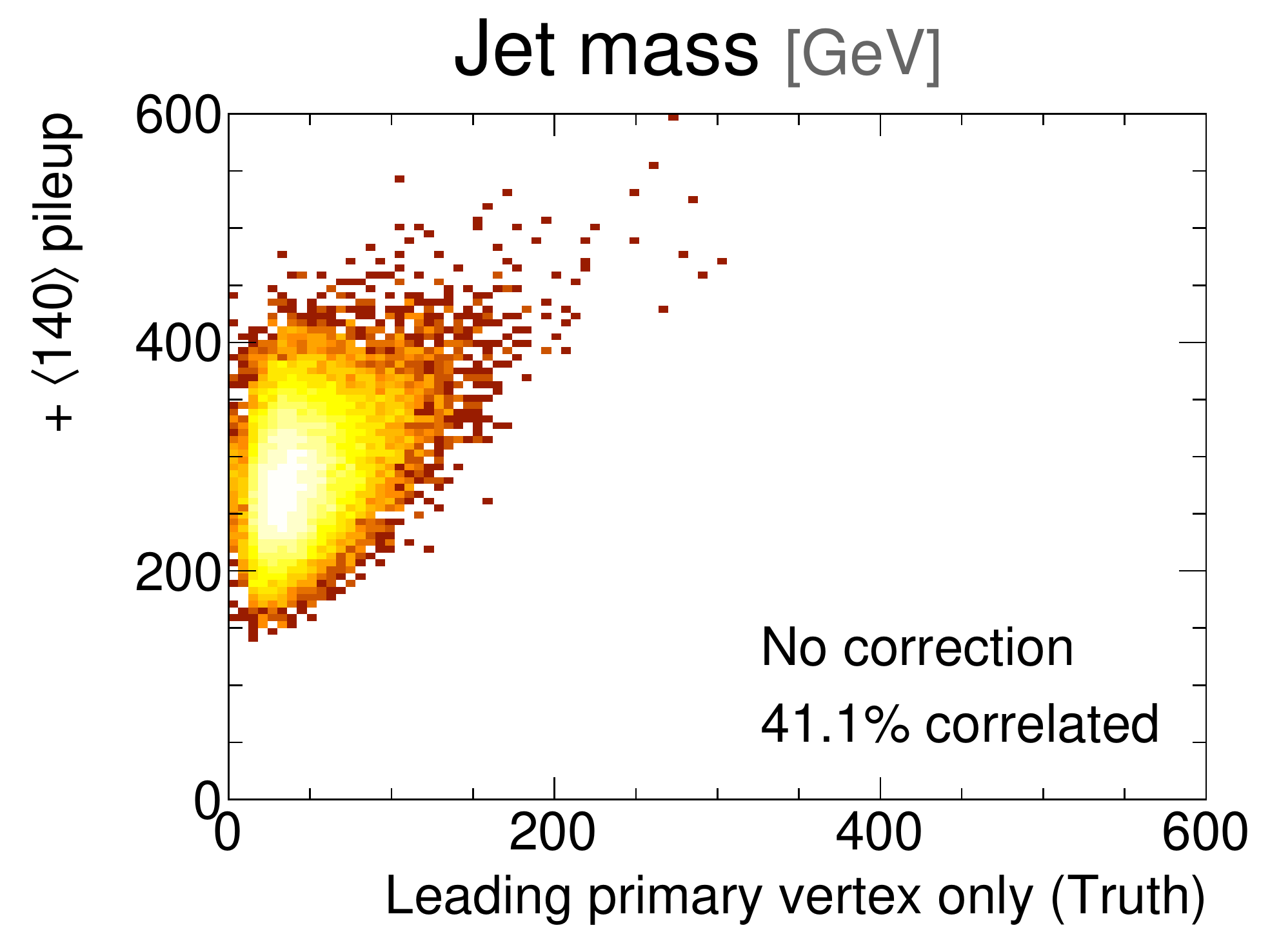}
  \includegraphics[width=0.40\textwidth]{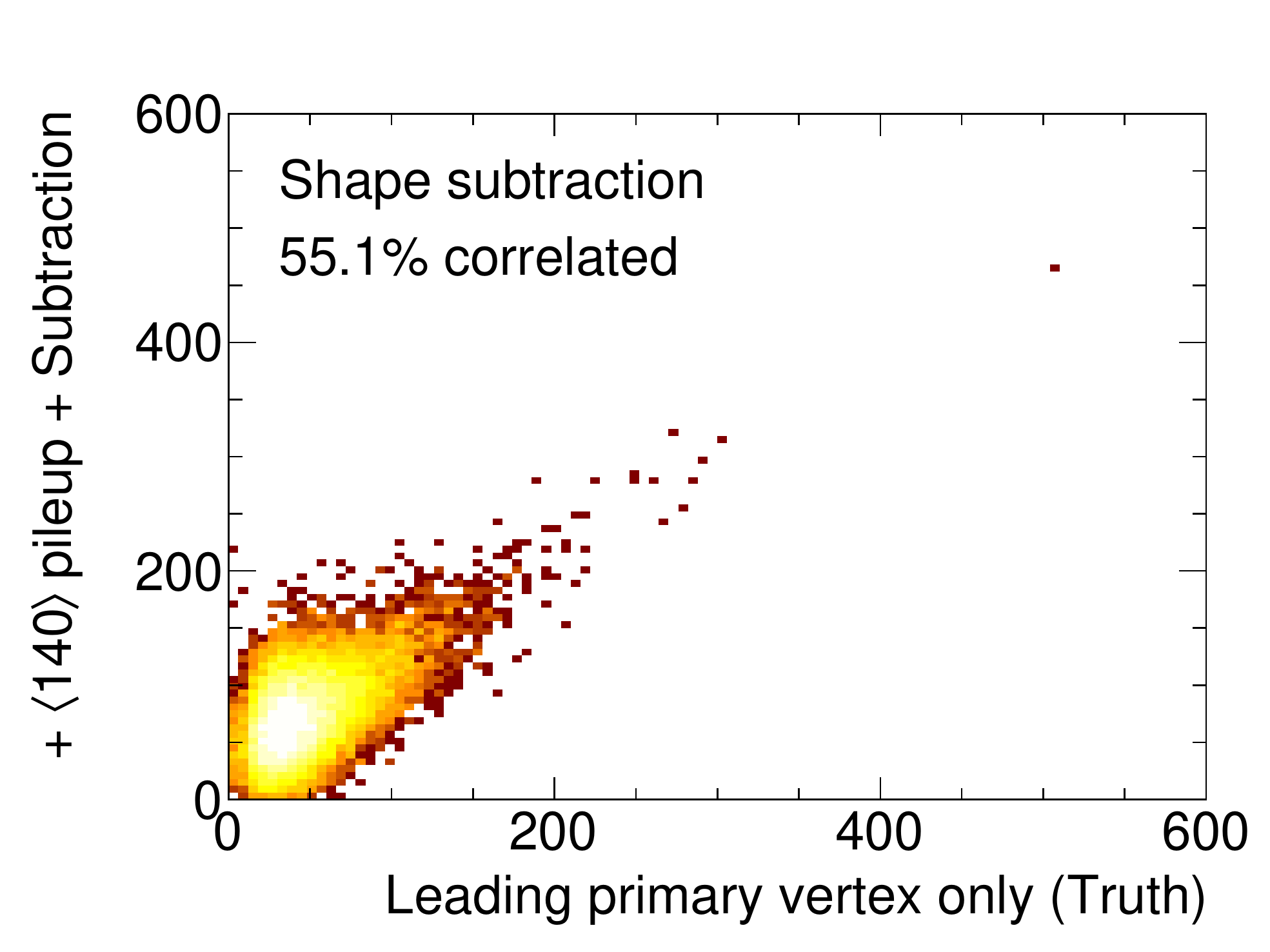}
  \includegraphics[width=0.40\textwidth]{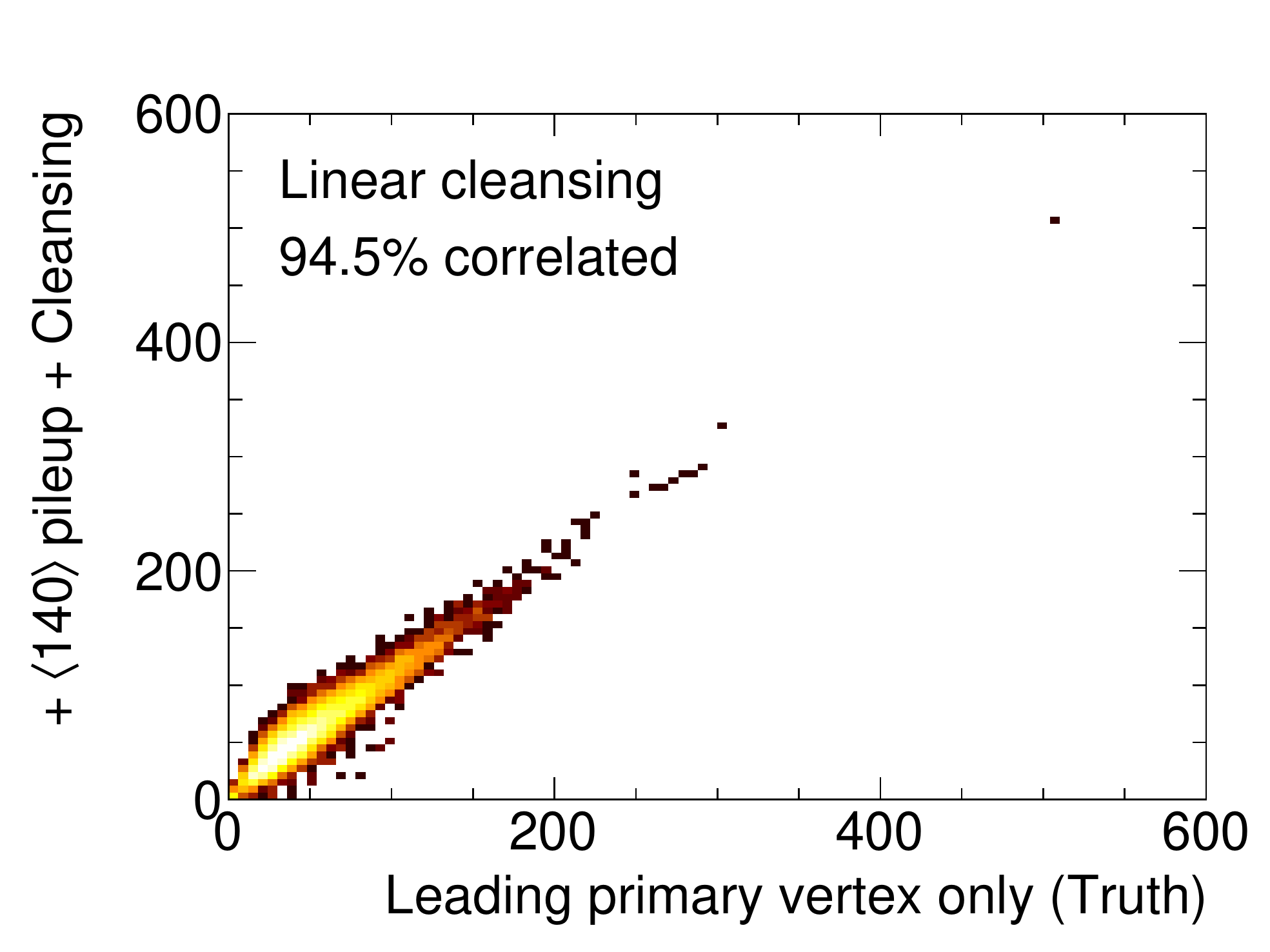}
  \caption{Correlations for jet mass, a substructure variable, are shown between between events with 140 pileup interactions, corrected via subtraction or cleansing, and the truth version of the same events, with pileup explicitly removed.  The top row shows the uncorrected correlations, the middle row demonstrates the performance of~\cite{Soyez:2012hv}, and the bottom row shows the performance of the linear cleansing method described here.}
  \vspace{-3ex}
  \label{fig:jet2D}
\end{figure}

Despite the success of these varied techniques, pileup is not a solved problem. 
None of the above methods is powerful enough to fully alleviate pileup's effects once $\npv\gtrsim{\cal O}(100+)$. This can be seen by comparing the top and bottom panels of Figs.~\ref{fig:dijetmass} and \ref{fig:jetmass}.  These figures show the results of various cleansing and subtraction techniques on a dijet mass resonance distribution and a jet mass distribution (see Sect.~\ref{sec:results} for a description of the dijet resonance used).  While with moderate pileup most methods work well, at higher levels their performance deteriorates.
The deterioration can also be seen in the 2D distribution of an observable with no pileup (using truth information)
and the observable after pileup is added and then subtracted/groomed. Such distributions are shown in Figs.~\ref{fig:dijet2D} and \ref{fig:jet2D}.
In addition, the assumption made by subtraction, that pileup is uniformly distributed over an event is inherently more effective for kinematic observables ({\it e.g.} jet $p_T$) than for jet shape observables ({\it e.g.} jet mass, $N$-subjettiness) which are sensitive to the distribution of radiation within a jet.
Furthermore, shape subtraction is performed as a Taylor expansion in the pileup density which can become inaccurate for large values of the expansion parameter, $\rho$.
In this paper, we present a method we call \emph{jet cleansing} which works at high pileup, is observable independent
and is remarkably effective for both kinematic and shape variables.

A new element introduced with jet cleansing beyond current experimental techniques like CHS and JVF  takes inspiration from early successful jet substructure techniques~\cite{Butterworth:2008iy,Cacciari:2008gd,Kaplan:2008ie,Krohn:2009th}. These
methods demonstrated the power of reclustering a large $R$ jet into jets of smaller $R$ and have been validated in data.
We find similarly that pileup removal can be much more effective if done on subjets with $R_\text{sub}=0.2$ or $R_\text{sub}=0.3$ rather than on full jets. Cleansing attempts to tailor the degree
of energy rescaling within a jet based on locally measured levels of charged and neutral particles.

\section{The Algorithm}
\label{sec:algorithm}

To produce the inputs to our algorithm, without access to full detector simulation, we make the following approximations and assumptions.
We discard all charged particles with $p_T < 500$ MeV. We then aggregate the remaining particles into $\Delta \eta \times \Delta \phi = 0.1\times0.1$ ``calorimeter cells'',
discarding any cells with $E < 1$ GeV. These calorimeter cells are then clustered into subjets of size $R_\text{sub}$.
 We assume the charged particles can all be tagged as either coming from the leading vertex or not,
and we associate them to the nearest calorimeter cell. The input to cleansing is therefore three numbers per subjet: the total transverse momentum, {\blue $p_T^{\text{tot}}$},
 the $p_T$ in charged particles
from the leading vertex, {\red $p_T^{\text{C,LV}}$}, and the $p_T$ from charged particles from pileup, {\red $p_T^{\text{C,PU}}$}. 
Jet cleansing aims to best extract the total momentum from the leading vertex only, {\green $p_\mu^{\text{LV}}$}, using these three inputs to rescale the four-vector constituents of the measured subjet ${\blue p_\mu^{\text{tot}}}$.

We propose three methods of varying sophistication with which {\green $p_\mu^{\text{LV}}$}
 can be guessed. Before explaining them, it is helpful to define 
$\gamma_0 \equiv {\red {p_T^{\text{C,PU}}}}/{\blue {p_T^{\text{PU}}}}$ and 
$\gamma_1 \equiv {\red {p_T^{\text{C,LV}}}}/{\blue {p_T^{\text{LV}}}}$. While we do not know $\gamma_0$ or $\gamma_1$ for any particular subjet,
they are constrained by
\begin{equation}
{\blue p_T^{\text{tot}}} =\frac{{\red p_T^{\text{C,PU}}}}{\gamma_0} +\frac{{\red p_T^{\text{C,LV}}}}{\gamma_1}.
\label{constraint}
\end{equation}

 The first method, which we call {\bf JVF cleansing} simply assumes $\gamma_0=\gamma_1$. This is the assumption that the charged-to-neutral ratio
is the same for pileup component and hard scatter component of jets. The result is that 
\begin{equation}
{\green p_\mu^{\text{LV}}}= {\blue p_\mu^{\text{tot}}} \times\frac{  {\red p_T^{\text{C,LV}}}  }{ {\red p_T^{\text{C,LV}}} + {\red p_T^{\text{C,PU}}} }.
\end{equation}
JVF cleansing is  similar to methods ATLAS has used (at the jet level). However, while effective, JVF
cleansing omits two important effects. First, 
there are large fluctuations in both $\gamma_0$ and $\gamma_1$ from subjet to subjet.  The other problem is that the expected values of $\gamma_0$
and $\gamma_1$ are not the same. The difference is largely due the fact that detector resolution treats soft and hard particles, and charged and neutral particles differently. 

\begin{figure}[t]
  \centering
  \includegraphics[width=0.40\textwidth]{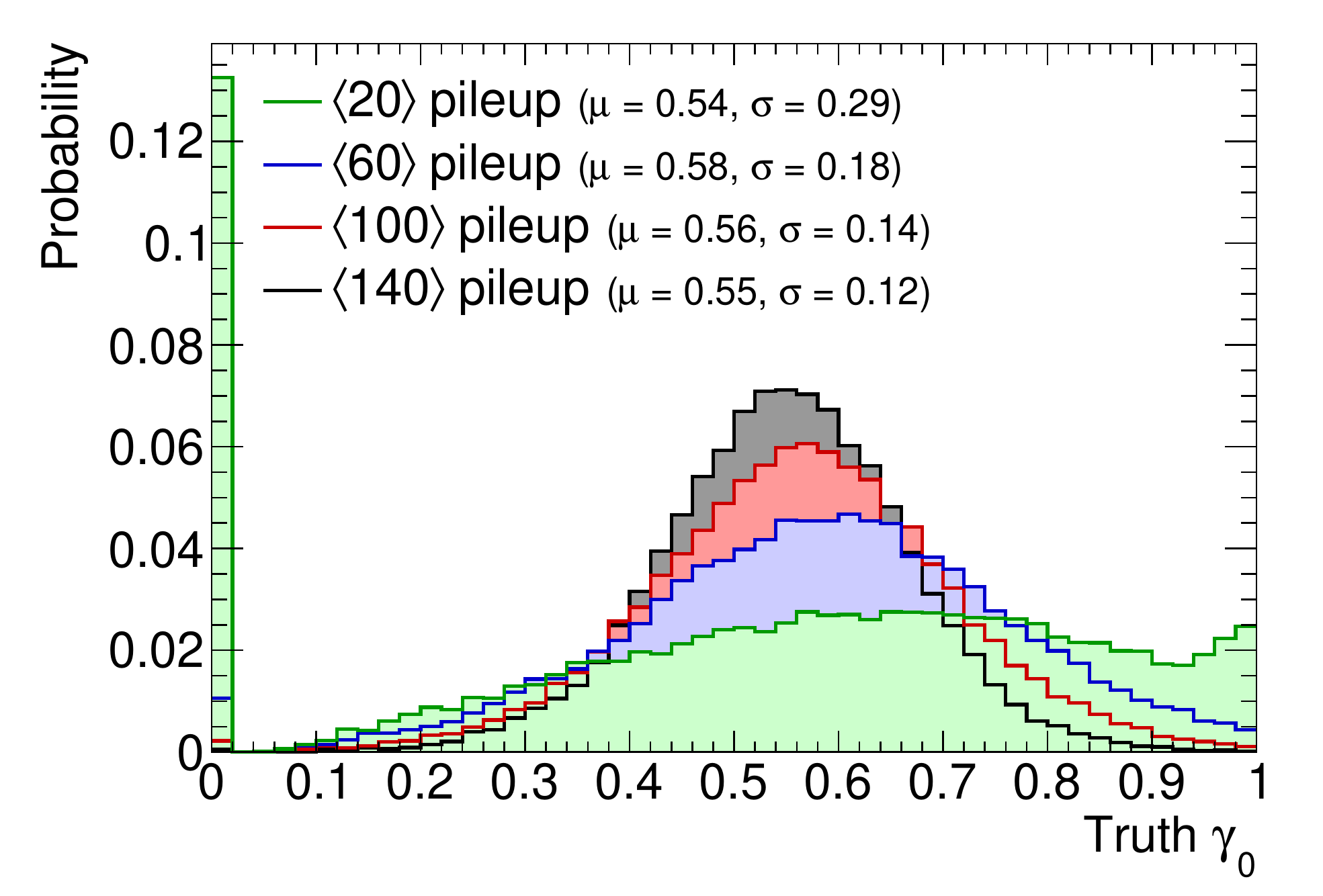}
  \includegraphics[width=0.40\textwidth]{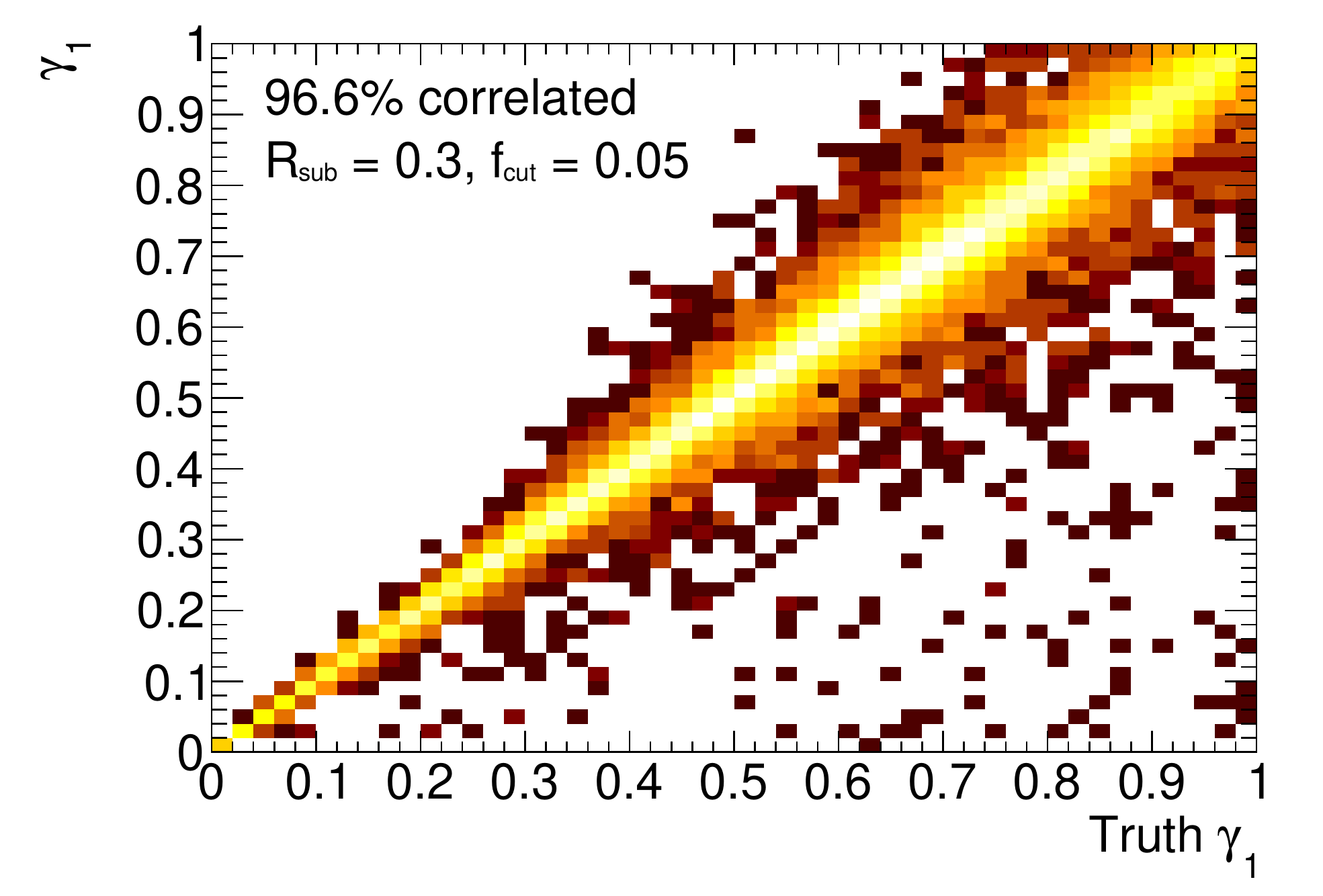}
  \caption{Top: the distribution of $\gamma_0$, the charged to total $p_T$ ratio in pileup, for various average number of pileup interactions.  Bottom: the correlation between the true value of $\gamma_1$, the charged to total $p_T$ ratio coming from the leading vertex, with its approximation using Eq.~\eqref{eq:solveg1}.
  \label{fig:g0}} 
\end{figure}

To improve on JVF cleansing, we observe that the $\gamma_0$ distribution is determined by fragmentation following many independent secondary collisions, while $\gamma_1$ is largely due to the fragmentation of a single hard parton. Thus, the fluctuations of $\gamma_0$ around its mean should
decrease with $\npv$, while the fluctuations of $\gamma_1$ are $\npv$-independent. This can be seen in Fig.~\ref{fig:g0}, which
shows the $\gamma_0$ distribution
for events with no leading vertex
 for various values of $\npv$. So an alternative to
JVF cleansing is to take $\gamma_0$ to be a constant, called $\overline{\gamma_0}$.
Based on Fig.~\ref{fig:g0}, we choose $\overline{\gamma_0}=0.55$. In fact, the distribution of $\gamma_0$ is sensitive
to how soft particles are handled. Ignoring detector effects it should be close to the isospin limit $\gamma_0\sim2/3$.
Experimentally, $\gamma_0$ can be determined from minimum bias events in data.

Taking $\gamma_0 = \overline{\gamma_0}$ for all subjets, we can then solve Eq.~\eqref{constraint} for $\gamma_1$. This gives
\begin{equation}
\gamma_1 = \frac{ {\red p_T^{\text{C,LV}}} }{  {\blue p_T^{\text{tot}}} - {\red p_T^{\text{C,PU}}}/  \overline{\gamma_0} } \label{eq:solveg1}
\end{equation}
The correlation of $\gamma_1$ from solving this equation with the truth-level $\gamma_1$ is shown in Fig.~\ref{fig:g0} at $\npv=140$. We find a 96.6\% correlation.
Using $\gamma_1$ to solve for ${\green p_\mu^{\text{LV}}}$ we get
\begin{equation}
{\green p_\mu^{\text{LV}}}= {\blue p_\mu^{\text{tot}}} \times\left (1- \frac{{\red p_T^{\text{C,PU}}}} {\overline{\gamma_0}\times {\blue p_T^{\text{tot}}}}\right).
\end{equation}
which is linear in  ${\red p_T^{\text{C,PU}}}$ and does not depend on ${\red p_T^{\text{C,LV}}}$ or the JVF. We call this method {\bf linear cleansing}\footnote{A version of linear cleansing applied to full jets (rather than subjets) used as a $p_T$ correction was discussed in~\cite{Salam:2011}.}.
As $\npv \to \infty$, the $\gamma_0$ distribution
as in Fig.~\ref{fig:g0} becomes sharper.
Thus, linear cleansing becomes more and more effective as pileup increases. 
 Even for moderate pileup, linear cleansing takes advantage of the fact that the stochastic nature of pileup makes
 the uncertainty on $\gamma_0$ less than on $\gamma_1$. Linear cleansing often yields an improvement over JVF cleansing and area/shape/charged hadron subtraction, as we quantify shortly\footnote{Occasionally, linear cleansing results in a negative rescaling.  When this happens we revert to JVF cleansing.  The frequency of JVF rescalings is a function of $\overline{\gamma_0}$, $\npv$, and the subjet's $p_T$.  For low $p_T$ subjets linear rescalings are used $\approx 60-80 \%$ of the time, while for subjets with $p_T \gtrsim 20~\gev$ linear rescalings are used $\approx 90-100 \%$ of the time.  Subjets with $p_T$ above $50~\gev$ are essentially all linearly rescaled.}.

In the third method, which we call {\bf Gaussian cleansing}, the $\gamma_1$ and $\gamma_0$ distributions are approximated as truncated Gaussians:
\begin{equation}
\label{eq:maxl}
P(\gamma_0,\gamma_1)\propto\exp\left[-\frac{1}{2}\sum_{i=0,1}\left(\frac{\gamma_i-\overline{\gamma_i}}{\sigma_i}\right)^2\right]
\end{equation}
for $0\leq\gamma_i\leq1$ where $\overline{\gamma_i}$ and $\sigma_i$ are the mean and widths of the Gaussian approximations\footnote
{In what follows we will take 
$\overline{\gamma_{0}}=0.55$,
$\overline{\gamma_{1}}=0.62$,
$\sigma_{0}=0.15$
and 
$\sigma_{1}=0.22$, although we have seen that the results are not very sensitive to these choices.
}. We then find the values of $\gamma_0$ and $\gamma_1$ satisfying Eq.~\eqref{constraint} which maximize Eq.~\eqref{eq:maxl}.
This approximation requires four input parameters but for this one is rewarded with further increases in performance.

We have implemented these three methods in a Fastjet plugin which can be obtained at \url{http://jets.physics.harvard.edu/Cleansing} and as part of the Fastjet Contrib project, \url{http://fastjet.hepforge.org/contrib}.

\section{Results}
\label{sec:results}

Below we compare each of the three cleansing methods to subtraction and CHS, all with and without jet grooming.  The details of our implementation of subtraction and CHS are found in App.~\ref{sec:simulation}.  We find that cleansing naturally dovetails with filtering and trimming, which already employ subjets\footnote{Under some definitions the use of subjets is already considered trimming.  Cleansing does not distinguish between no trimming and trimming with $f_{\text{cut}}=0$.}.  Where grooming is applied we adopt the trimming procedure which supplements cleansing by applying a cut on the ratio $f$ of the subjet $p_T$ (after cleansing) to the total jet $p_T$.  Subjets with $f<f_{\text{cut}}$ are discarded.  

Our signal process comes from a color-singlet resonance with a mass of 500 GeV decaying into $q\bar q$ dijets,  while our background is from QCD dijet events all at $E_{\text{CM}} = 13$ TeV.  Jets are clustered using the anti-$k_T$~\cite{Cacciari:2008gp} algorithm with $R=1.0$~\footnote{We choose $R=1.0$ for simplicity, different procedures may have different optimal $R$ values.  However, we have seen that varying the choice of $R$ does not change out conclusions.} implemented in Fastjet v3.0.3~\cite{Cacciari:2011ma}.  Where we do apply jet cleansing we employ $R_{\text{sub}}=0.3$ subjets\footnote{In general we find smaller $R_{\text{sub}}$ offers marginal improvement.} and take $f_{\rm cut}=0.05$ where trimming is used.  Further technical details of the simulation are found in App.~\ref{sec:simulation}.  

\begin{figure}[t]
   \includegraphics[width=0.40\textwidth]{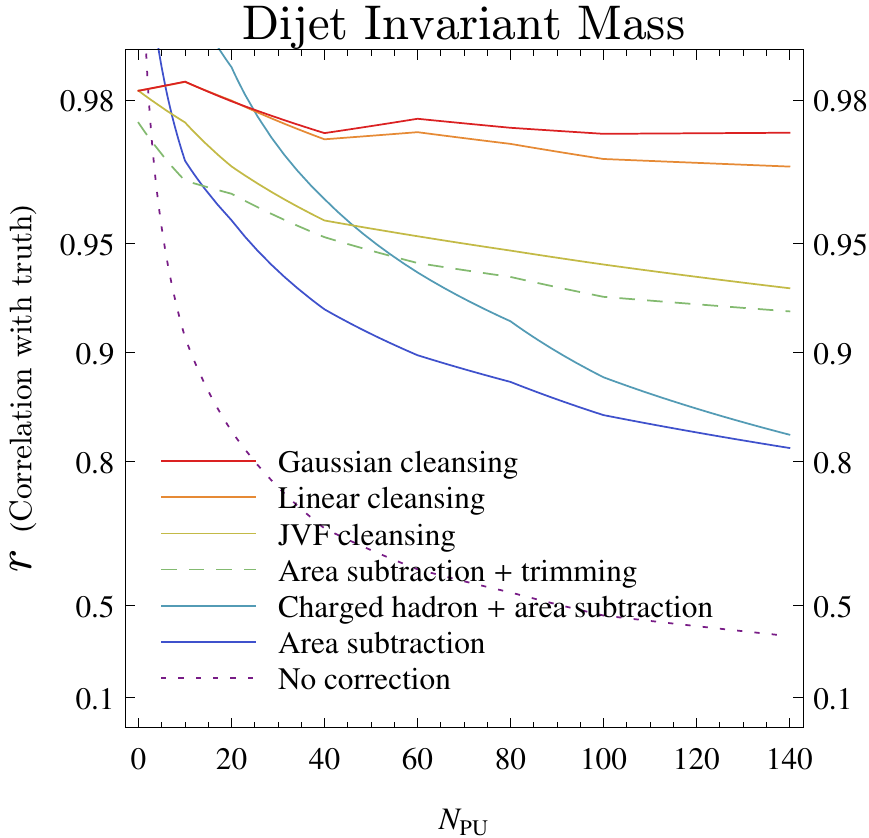}
   \includegraphics[width=0.40\textwidth]{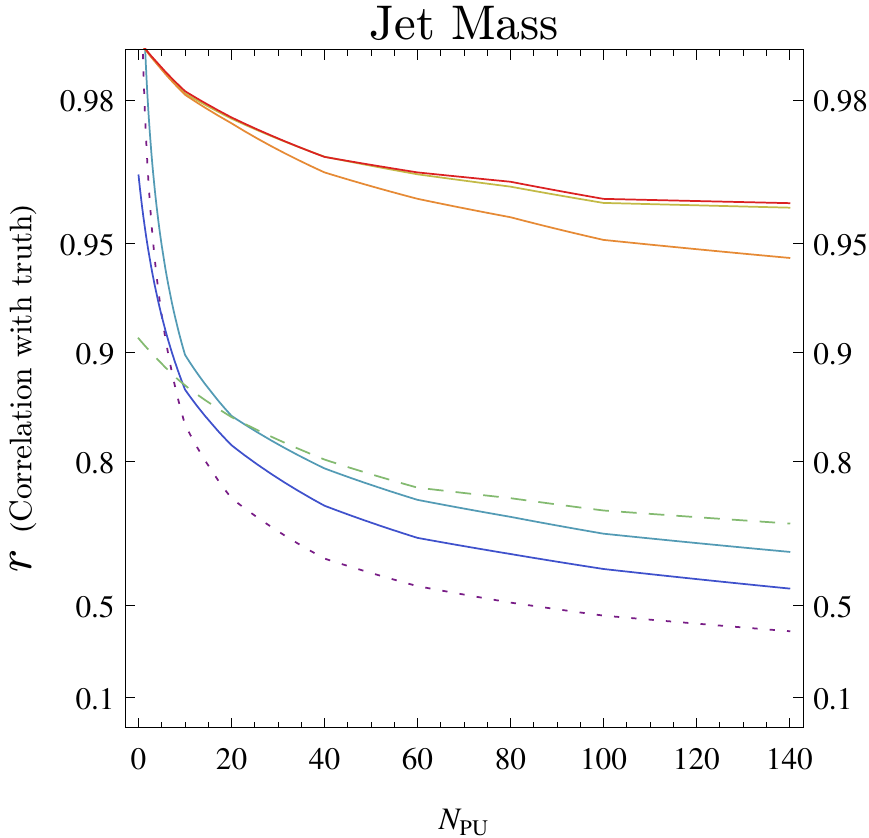}
\caption{Linear correlation coefficients as a function of pileup for the dijet invariant mass and the jet mass.  Higher values are better.
  \vspace{-3ex}
  \label{fig:CorrPU}
}
 \end{figure}

To test jet cleansing, we compare its performance to other methods in the reconstruction of both a kinematic variable, the dijet invariant mass, and a jet shape variable,
the jet mass. As measures of performance, we consider
significance as approximated by $S/\sqrt{B}$,
where $S$ and $B$ are the number of signal or background events in a 40 GeV window (the center of the window is floated separately to optimize significance for each method).
For signal events, we also compute the Pearson linear correlation coefficient $r$ between the observable
with and without pileup contamination.  The correlation coefficient is a useful measure here because the objective of cleansing is to restore the full representation of the jet.  While correlations can be sensitive to the tails of distributions a high correlation indicates the method is successfully returning the jet to its uncontaminated state.

Results for the dijet invariant mass and jet mass are presented in Tables~\ref{table:resultsD} and~\ref{table:resultsM} respectively.
Correlations for sample distributions are shown in Figs.~\ref{fig:dijet2D} and \ref{fig:jet2D} and the correlation coefficients as a function of $\npv$ are plotted in Fig.~\ref{fig:CorrPU}.  As one can see from the tables and plots, jet cleansing yields the best performance in every test case. Both area and shape subtraction can be improved by working at the subjet level, and
including a mild amount of trimming, yet even with these improvements cleansing still comes out ahead. Also, as mentioned above, cleansing is especially  effective at improving measurements of observables like jet mass which are more sensitive to the spatial distribution of radiation within a jet. 
We therefore expect cleansing also to work well on $N$-subjettiness~\cite{Thaler:2010tr,Kim:2010uj,Thaler:2011gf} which is especially sensitive to contamination.

That Gaussian cleansing tends to work better than linear cleansing is not surprising, since it is a more sophisticated algorithm. However,
Gaussian cleansing needs input about the $\gamma_1$ distribution which is related to the signal process. Although results are not that sensitive to the precise values
of widths and means of the Gaussians used as inputs, there could be some process/energy dependence if optimal performance is desired. In contrast, linear cleansing only
requires an estimate of $\overline{\gamma_0}$ which can be extracted from minimum bias data. 

Finally, it is worthwhile to note the role played by trimming.  As jet cleansing is designed to locally remove pileup it achieves the best correlations when used without trimming.  If one wishes to approximate the jet's pre-pileup state, cleansing alone is the best option.  Trimming offers improvement when the objective is to maximize $S/\sqrt{B}$.  This makes sense because trimming is known to be useful, even when applied on jets with no pileup, as it tends to remove underlying event and other soft contamination primarily leaving the final state radiation.  For subtraction and CHS, in contrast, at high levels of pileup trimming is important both for correlations and $S/\sqrt{B}$.

\begingroup
\begin{table}[t]
\begin{tabular}{|c|cc|cc|}
\hline
\multicolumn{5}{|c|}{{\bf Significance improvement}} \\
\hline
\multirow{2}{*}{Algorithm}  & \multicolumn{2}{c|}{$\npv=20$} &  \multicolumn{2}{c|}{$\npv=140$}\\
& plain~~ & trimmed & plain~~ & trimmed \\
 \hline
 \hline
  {\rm CH + area Sub.}    & 0.86 & 1.07 & 0.48 & 0.90 \\
  {\rm Area subtraction}  & 0.87 & 1.00 & 0.45 & 0.85 \\
  {\rm JVF cleansing}     & 0.93 & 1.06 & 0.82 & 0.81 \\
  {\rm Linear cleansing}  & 0.94 & 1.08 & 0.78 & 1.00 \\
  {\rm Gaussian cleansing}& 0.95 & 1.07 & 0.91 & 0.98 \\
 \hline
\end{tabular}
 \caption{The ratio  $S/\sqrt{B}$ for a variety of algorithms and levels of pileup, divided by
$S/\sqrt{B}$ from events with no pileup using plain anti-$k_T$ jets.  Larger values are better.  We estimate the statistical uncertainty on these numbers to be $\pm0.05$.}
 \label{table:resultsD}
\end{table}
\endgroup

\section{Conclusions and Outlook}
\label{sec:conclusion}

Jet mass has been calculated to high accuracy using perturbative QCD~\cite{Dasgupta:2012hg, Chien:2012ur, Jouttenus:2013hs}, and measured in 7 TeV LHC data~\cite{ATLAS:2012am,Chatrchyan:2013rla}.  A direct comparison between these calculations and the data has been limited by the contamination of pileup. Since the improvement in pileup removal of cleansing over shape subtraction for jet mass are substantial, there is now hope that precision QCD jet shape (and jet distribution) calculations can be productively compared to data from the high luminosity LHC runs.

While jet cleansing works extremely well at high pileup, it is not perfect. It would be interesting to explore whether it could be improved by combining it with jet area subtraction, or by exploiting the probabilistic approach as in the Qjets paradigm~\cite{Chien:2013kca, Ellis:2012sn, Kahawala:2013sba}. It would also be interesting to see if cleansing can reduce the uncertainty on missing energy measurements.  Finally, a note of caution -- jet energy uncertainties~\cite{Aad:2012ag,CMS-PAS-JME-10-014} may ultimately limit the performance of jet cleansing.  However, given the potential improvements provided by cleansing over current methods, especially in the reconstruction of jet shapes, it is likely that cleansing will still be useful when full detector effects are included.

\noindent
{\it Note added:} Shortly after the preprint was posted, ATLAS demonstrated cleansing works in its full detector simulation at up to pileup levels of $\npv = 40$~\cite{ATL-PHYS-PUB-2014-001}.

\begingroup
\begin{table}[t]
\begin{tabular}{|c|cc|cc|}\hline
\multicolumn{5}{|c|}{{\bf Distance correlation (\%)}} \\
\hline
\multirow{2}{*}{Algorithm}  & \multicolumn{2}{c|}{Jet mass} & \multicolumn{2}{c|}{Dijet mass}\\
&$\npv = 20$&$~~140~$&$\npv = 20$&$~~140~$\\
 \hline
 \hline
  {\rm CH + area Sub.}    & 20  & 37  & 0.9 & 13  \\
  {\rm Shape/area Sub.}   & 18  & 45  & 2.9 & 15  \\
  {\rm JVF cleansing}     & 2.3 & 4.0 & 1.6 & 3.6 \\
  {\rm Linear cleansing}  & 2.3 & 5.5 & 1.1 & 1.7 \\
  {\rm Gaussian cleansing}& 2.2 & 3.9 & 1.1 & 1.3 \\
 \hline
\end{tabular}
 \caption{The distance correlation, $d=1-r$ in percent, where $r$ is the linear correlation coefficient between jet mass
 or dijet mass as measured on pileup-free samples and samples with various levels of pileup.  Smaller values are better --- they indicate higher correlation.}
 \label{table:resultsM}
\end{table}
\endgroup

\vspace{-1em}
\begin{acknowledgments}
The authors would like to especially thank A. Schwartzman for informative discussions of systematics in jet energy measurements and detailed comments on the manuscript.  The authors would also like to thank Y.-T. Chien, J. Dolen, S. Ellis, M. Freytsis, P. Harris, J. Huth, T. Lin, P. Loch, D. Lopez Mateos, D. W. Miller, S. Rappoccio, G. Salam, M. Swiatlowski, N. Tran, W. Waalewijn, and J. Walsh for useful discussions.  ML is supported by an NSERC PGS D fellowship, DK is supported by a Simons postdoctoral fellowship,  MDS is supported by DOE grant DE-SC003916, and LTW is supported by DOE grant DE-SC0003930.  Computations for this paper were performed on the Hypnotoad cluster supported by PSD Computing at the University of Chicago.
\end{acknowledgments}

\appendix
\section{Monte Carlo Details}
\label{sec:simulation}

The signal sample used in this study was a color-singlet scalar resonance with mass $m_\phi=500~\gev$ decaying to light quarks.  Signal events were generated at matrix-element level using Madgraph5 v1.5.8~\cite{Alwall:2011uj} requiring that the quarks have $p_T>95~\gev$.  Pythia v8.176~\cite{Sjostrand:2007gs}, tune 4C, was used to shower and hadronize events.  The background sample used was hard QCD dijet events as implemented in Pythia using a phase space cut requiring partons with $p_T>95~\gev$.  To simulate pileup events, for each event $i$ we overlay $n_i$ soft QCD events drawn from a Poisson distribution with mean $\npv$.  The soft QCD events are generated in Pythia.  All samples are generated at $E_{\text{CM}}=13~\tev$.

Jets are clustered from the full event, including the hard scatter and pileup, using the anti-$k_T$ algorithm~\cite{Cacciari:2008gp} with $R=1.0$ as implemented in Fastjet v3.0.3~\cite{Cacciari:2011ma}.  For each event, the two hardest jets are kept provided they have $p_T>150~\gev$ and $|\eta|<2.5$.  These jets are used in the jet mass distributions and events with both of the two hardest jets passing these cuts are used in the dijet mass distributions.  Where trimming is used we employ $R_{\text{sub}}=0.3$ subjets and take $f_{\text{cut}}=0.05$.

In correlations, the groomed/subtracted/cleansed jet is compared against the ``truth'' jet.  The truth jet is constructed by removing all of the four-vectors originating from pileup leaving only contributions from the underlying hard scatter.  In cases where particles from pileup and the hard scatter fall into the same cell, the cell is kept massless but rescaled to its hard scatter value by multiplying the four-vector by $E_{\text{cell,LV}} / E_{\text{cell}}$.

\section{Subtraction Methods}
\label{sec:submethods}

\noindent
{\it Area subtraction}:  Jets corrected by area subtraction via $p^\mu_{\text{corr}} = p^\mu - \rho A^\mu$, where $\rho$ is a measure of the event density and $A^\mu$ is the four-vector area.  To compute $\rho$, the event is tiled in $k_T$ jets with $R=0.5$ up to $|\eta|<4.0$ and $\rho$ is taken as the median of the $p_T/\text{area}$ distribution.  This is done using the native Fastjet implementation of {\tt JetMedianBackgroundEstimator}.  For each event we use the global value of $\rho$ and do not include rapidity dependence for simplicity.  We have checked that the improvements from including rapidity dependence are small and do not affect any of the conclusions.  The area of each jet is computed using the Fastjet implementation of areas.  We use the jet area from the full event which includes the effect of pileup.

\vspace{1em}
\noindent
{\it Charged hadron subtraction}:  Our implementation of charged hadron subtraction proceeds as follows.  First all four-vectors that come from charged pileup are subtracted from the jet.  Next, we compute $\rho_{\text{neutral}}$ for the event, using the same method and parameters as above, but only including neutral particles.  Finally $\rho_{\text{neutral}} A^\mu$ is subtracted from the jet, with charged pileup already removed, where $A^\mu$ is the area found from the full event.

\bibliography{cleansing}
\end{document}